# Cross Layer Adaptation for QoS in WSN

Sukumar Nandi and Aditya Yadav

Department of Computer Science and Engineering
IIT Guwahati

**Abstract.** In this paper, we propose QoS aware MAC protocol for Wireless Sensor Networks and its cross layer extension to network layer for providing QoS in delay sensitive WSN scenarios. In WSNs, there can be two types of traffic one is event driven traffic which requires immediate attention and another is periodic reporting. Event driven traffic is classified as Class I(delay sensitive) traffic and periodic reporting is classified as Class II(Best Effort) Traffic. MAC layer adaptation can take place in terms of (i) Dynamic contention window adjustment per class, (ii) Reducing the delay suffered by difference in Sleep schedules(DSS) of communicating nodes by dynamically adjusting Duty Cycle based on Utilization and DSS delay of class I traffic, (iii) Different DIFS (DCF Inter Frame Spacing) per class, (iv) Adjusting all the three schemes proposed above simultaneously. Cross layer extension is also proposed, in which MAC layer uses network layer's next hop information for better adaptation of duty cycle based on DSS delay. Routing protocols can utilize MAC layer parameter DSS delay to select the routes which offer least DSS delay latency, there by minimizing the overall end-to-end delay.

## 1 Introduction

WSNs are deployed for critical monitoring applications, some of these applications are delay-sensitive and have stringent requirement on end-to-end latencies. Apart from this delay-sensitive traffic, they also support periodic reporting of environment to their base stations. WSNs deployment for forest fire monitoring is one of the type of applications that we are emphasizing on. This calls for QoS specific mechanisms to be in place for supporting the application requirements. There has been a lot of research and development carried out in architecture, protocol design, energy saving and location in WSNs, but only a few studies have been done regarding network efficiency (i.e. Quality of Service QoS) in WSNs.

We consider SMAC[1], as one of the widely accepted MAC layer implementations in WSNs and use it as our base MAC protocol for implementing the QoS features at MAC level. SMAC pays no attention to latency and end to end delay, and nodes form virtual clusters in terms of their common sleep-wakeup schedule to reduce control overhead. For supporting delay-sensitive traffic and periodic reporting for above mentioned application requirements, our prototype defines two classes of traffic classI(immediate attention) and classII(periodic reporting) respectively. In our prototype, we propose to adjust MAC level parameters Contention Window[7], DIFS[7], Sleep schedules of nodes[1],[2] to provide QoS for the classes of traffic specified. In [9] MAC level parameters are adjusted for QoS guarantees, In [2], duty cycle is being adapted, but efficient QoS classes specific way of adapting duty cycle in case of SMAC[1], and adapting MAC level parameters Contention Window and DCF- IFS in case of Sensor Networks is our contribution, that this paper proposes.

MAC layer is responsible for scheduling and allocation of the shared wireless channel, which eventually determines the link level QoS parameters namely MAC delay. To maintain the per class service guarantees in dynamic environment, MAC layer is made adaptive to current network conditions. The proposed dynamic adaptation of the behavior of MAC layer is done by, (i) dynamically varying the contention window based on the class of traffic, (ii)Reducing the delay suffered by Difference in Sleep Schedules(DSS) of communicating nodes by dynamically adjusting Duty Cycle based on Utilization and DSS delay of class I traffic, (iii) By differentiating the DCF Inter Frame Spacing for different class of Traffic, (iv) Combining all the three proposed schemes for MAC Layer Adaptation. [12], reports the premilimnary version of the work presented in this paper.

Cross layer extension to current MAC layer adaptation is also proposed. In this, MAC layer can be adjusted to use network layer next hop information for adjusting the duty cycle more efficiently, next hop information is broadcasted with the SYNC packets used in SMAC[1]. So that only those nodes which are next hops of many other nodes should increase their duty cycle based on utilization and DSS delay. In second cross layer mechanism, we focus on routing protocols, in this link costs can include DSS delay parameter of underlying MAC layer into consideration while route calculation. So, only those routes are chosen which offer least overall DSS delay and in turn least end-to-end delay.

The rest of the paper is organized as follows. In section 2 we briefly discuss about the related work. Section 3 describes our proposed scheme for MAC layer adaptation. Section 4 discusses about the cross layer extensions. Section 5 presents the simulated results to show the the efficiency of our proposed scheme. Section 6 concludes the paper.

## 2  RELATED WORK

Some of the works in improving end-to-end latency by duty cycle adaptation are DSMAC[2] and UMAC[6].

In DSMAC[2], when duty cycle of a node is changed, for maintaining same sleep schedule of this node with its neighbors, duty cycle is always doubled or halved, so that neighbors can follow the same schedule and still communicate with duty cycle changed node. The only difference in the new duty cycle node now either wakes up more or less frequently than its neighbors. In DSMAC [2], they take into account the delay faced by the sending nodes to a receiving node for changing the duty cycle, but change of twice or half, increases the energy consumption in Sensor networks. Moreover, it doesn't consider the adaptation for different types of traffic in WSNs.

In another scheme, UMAC [6] duty cycle is changed dynamically and its not always doubled or halved, but its changed according to Utilization of the node, it pays no attention to delay suffered by one hop neighbors of a node in sending data. It also neglects the different service requirements for different class of traffic.

There has been some works done on providing QoS in WSNs at different layers, mainly on Network and MAC layer. One of such work is, Energy aware QoS Routing. In [3], the authors propose a QoS-aware protocol for realtime traffic generated by WSNs, consisting of image sensors. This protocol implements a priority system that divides the traffic flows in two classes: best effort and realtime. All nodes use two queues, one for each traffic class. This way, different kinds of services can be provided to these types of traffic.

Another MAC level contribution for QoS in WSNs is B-MAC [4]. It stands out for its design and implementation simplicity, which has an immediate effect in memory size occupation and power saving. B-MAC does not implement any specific QoS mechanism; however, this fact is compensated by its good design. Some parts of this design are addressed to improve the efficiency for avoiding collisions, efficiency in the channel occupation at low and high data rates, the tolerance to changeable environments, or the good scalability properties.

In [5], author proposes a traffic-aware MAC protocol which dynamically adjusts the duty cycle adapting to the traffic load. Adaptive scheme operates on a tree topology, and nodes wake up only for the time measured for successful transmissions. By adjusting the duty cycle, it can prevent packet drops and save energy.

In [11], author proposes a cross layer scheme involving MAC layer and Network layer for energy efficient usage of wireless sensor networks. In this nodes, along with the RTS also include the information of final destination and subsequent CTS includes the information of upcoming nodes address, so that when the NAV timer of the nodes expire only that node wakes up and rest other sleep, thereby resulting in energy efficient usage.

## 3  Dynamic MAC Layer Adaptation For QoS in WSNs

In our framework, MAC layer is designed to adapt their behavior based on the dynamic network conditions(which can be obtained through continuous monitoring) and service quality requirements of the admitted traffic.

### 3.1  Estimation of MAC delay

The measurement technique for MAC delay is very simple. A node computes the MAC delay(d) by subtracting the time ($t_r$), that a packet is passed to the MAC layer from the time ($t_s$), an the packet is actually sent onto the link. Here $d^i$ and $d_{avg}^{i-1}$ are measured MAC delay and previously stored average MAC delay for a service class. $\eta$ is a positive constant, which determines how much effect the previously stored average MAC delay have on the current average MAC delay. The contention window rules for our two service classes are given below.

$$d_{avg}^i = (1 - \eta) * d^i + \eta * d_{avg}^{i-1} \qquad (1)$$

### 3.2 Contention Window Adaptation

One of the schemes that we propose for adapting MAC Layer for providing QoS in WSNs is Dynamic Contention Window adaptation for different class of traffic. The contention window parameters namely CWmin and CWmax provides intra node service differentiation among different class of traffic. The different service classes are assigned a non overlapping ranges of contention window in default setting. These default contention window ranges are adjusted by the CW Adaptation based on dynamic network conditions and required service level of a QoS class. To provide service differentiation across different classes of traffic, the non overlapping contention window ranges are maintained while performing contention window adaptation.

**Class I(Delay sensitive service)** This class of traffic corresponds to event driven traffic(immediate attention) in WSNs. We have assumed that each node ensure a maximum MAC Delay of $D_I$ for traffic of class I, and try to maintain it by periodically monitoring the observed delay in that class and adjusting the CW range accordingly. In this case, based on the per hop delay requirements of the class I, given as $D_I$, and the current MAC delay, D, measured in the node for Class I traffic, $CWmax$ is adjusted as shown in Algorithm 1. In the algorithm, $CWmax^{prev}$ represents the value of CWmax for class I before applying CW adaptation, and $CWmax$ is the new value of CWmax obtained by applying the dynamic adaptation of contention window, due to the difference between expected maximum delay, $D_I$, and the current measured delay, D, for class I. To avoid the possibility of unnecessary fluctuation in setting of CWmax, a threshold value namely Contention Window Threshold, $CWthresh_I$ is used in the algorithm. Here $\alpha_I$ is a small positive constant, that should be selected appropriately to enable faster adaptation of CWmax to prevailing network condition.

**Class II(Best Effort Service)** This class of traffic corresponds to periodic reporting in WSNs. Traffic in this class has no delay guarantee requirements. Contention window adaptation for this class of service is required to properly utilize the available resources in the network, without degrading the service quality of other high quality service class. Because the contention window of Class II traffic, should not degrade the performance of ongoing higher priority traffic, a checking is performed such that the CWmin value of Best Effort traffic will not be smaller than CWmax of the other high priority traffic.

**Algorithm 1** Procedure for Contention Window Adjustment of Class I

```
{CWmax is calculated based on Mac delay}
CWmax := CWmax^prev * (1 − α_I * (D − D_I) ÷ D)
if (abs(CWmax − CWmax^prev) < CWthresh_I) then
    CWmax := CWmax^prev
    return
end if
if (CWmax < CWmax^prev) then
    CWmax := max(CWmax, CWmax^def)
else
    if (CWmax > CWmax^prev) then
        CWmax := min(CWmax, CWmax^max)
    end if
end if
```

For class I, CWmin(default) value is 7, CWmax(default) is 15 and CWmax(max) is 31. For class II, CWmin(default) value is 32, CWmax(default) is 63 and CWmax(max) is 63.

### 3.3 Duty Cycle Adaptation

Second scheme, that we propose for adapting MAC Layer for providing QoS in WSNs is Dynamic Duty Cycle change based on Utilization of a node and one hop delay to the receiving node for classI traffic. In our proposed scheme of duty cycle adaptation, we change duty cycle taking into account both the criteria of Utilization and delay suffered by one hop neighbors. We change the duty cycle by appropriate percentage, by taking into account two factors mentioned above, as specified in Algorithm 2.

In Wireless Sensor Networks, the MAC layer protocol SMAC[1], we noticed that main reason behind delays in WSNs is the different sleep schedules followed by the nodes. In SMAC[1], Nodes form virtual clusters based on common sleep schedules to reduce control overhead and enable traffic-adaptive wake-up. So when data is sent from source to sink, at many nodes it has to go from one virtual clusters to another virtual cluster, and the bordering node follows both the sleep schedules. This difference in sleep schedule is incorporated in total end-to-end delay.

For the Class I(delay sensitive) traffic, this delay is undesirable. One of the solutions to this problem that we propose, is to dynamically vary the duty cycle of the nodes which are receiving more of Class I traffic. Suppose some Node 1 tries to transmit some class I traffic to Node 2, due to difference in sleep schedule, it will incur a delay, If many nodes wants to transmit to Node 2, almost all will be incurring this delay, which will increase the overall end to end delay of class I traffic. So we propose to increase the duty cycle of Node 2 based on utilization of the node and delay information received in the data frames by the sending nodes.

Estimation of Difference in Sleep Schedules(DSS) Delay The measurement technique for Difference in Sleep Schedules Delay is very simple. A transmitting node computes this delay(s), by subtracting the time ($ts_s$) that a packet is passed to the MAC layer from the time ($ts_c$) it starts carrier sensing for sending the packet. Then it includes this delay ($ts_s - ts_c$) in the MAC frame header and sends the frame. Receiving node then extracts this delay as $s^i$, and calculates the average delay $s^i_{avg}$. Here $s^i$ and $s^{i-1}_{avg}$ are measured DSS delay, and previously stored average DSS delay for a service class. $\zeta$ is a positive constant, which determines how much effect the previously stored average DSS delay have on the current average DSS delay.

$$s^i_{avg} = (1 - \zeta) * s^i + \zeta * s^{i-1}_{avg} \qquad (2)$$

We vary the duty cycle at the time of synchronization, when in SMAC it broadcasts the SYNC packets for the neighbor discovery, and each node also periodically broadcasts the SYNC packets so that its synchronized with its neighboring nodes. So, when a node changes its duty cycle, it broadcasts in its SYNC packets its new updated time before it goes to sleep, and in this way changing the duty cycle doesn't desynchronize the nodes. Moreover by this, we reduce the difference in sleeping schedules of all other nodes with this node, which has just updated its duty Cycle.

For changing the duty cycle by appropriate amount as calculated in Algorithm 2, duty cycle of a node should be updated at the time of sending SYNC packets, so that its not desynchronized with other nodes as we are not changing Duty Cycle by double or half. A node may be in many different sleep synchronized virtual clusters, it is following many common sleep schedules. So when duty cycle is changed it should be changed in all the schedules and all the nodes following those schedules which are at one hop distance to current node, should be informed by SYNC packets.

SYNC packets are sent after some periods defined in SMAC[1] as SYNCPERIOD. We implement this scheme by choosing the primary or first schedule followed by node, as the schedule to change the duty cycle. Algorithm 2 , is performed when SYNC has to be broadcasted in this schedule. Suppose duty cycle needs to be changed, then we broadcast the SYNC in this schedule to indicate the change in duty cycle, and for other schedules which till now also has some non-zero periods left before they transmit SYNC packets, we make number of periods left to zero, so that neighboring nodes following those schedules can perceive the change in duty Cycle of the SYNC packet sending node, and by this way nodes are not desynchronized even after varying duty cycle dynamically.

Duty Cycle Adaptation for Class I(Delay sensitive service): We have assumed that each node ensure a maximum DSS Delay of $S_I$ for traffic of class I, and try to maintain it by periodically monitoring the observed delay in that class, and adjusting the Duty Cycle accordingly based on Utilization. In this case, based on the per hop delay requirements of the class I, given as $S_I$, and the current DSS delay, S, measured in the node for Class I traffic, Duty Cycle(DC) is adjusted as shown in Algorithm 2. In the algorithm, $DC^{prev}$ represents the value of DC for class I before applying Duty Cycle adaptation, and DC is the new value of Duty Cycle obtained by applying the dynamic adaptation of Duty Cycle, due to the difference between expected maximum delay, $S_I$, and the current measured delay, S, for class I. To avoid the possibility of unnecessary fluctuation in setting of DC, a threshold value namely Duty Cycle Threshold, DC thresh is used in the algorithm. Here, $U_{min}$ is min Utilization to change DC, $DC_U$ is permissible DC calculated from Utilization, ρ is ratio of classI and class II packets.

For classI $DC_{min}$ is 30, and $DC_{max}$ is 60, and $DC_{default}$ for classI and classII is 30. DC thresh can be chosen as 5% $DC^{prev}$. We don't want to fluctuate on very small changes. $U_{min}$ is dependent on the traffic rate, according to our traffic rate we took $U_{min}$ as 10%. $ρ_{min}$ is application dependent, depending on how delay sensitive is application. We took $ρ_{min}$ to be 30%. $U_{min}$ and $ρ_{min}$ these are traffic and application dependent parameters, that depends on functionality performed by WSNs. $U_{prev}$ is the previous value of Utilization, we increase the DC in proportion to increase in Utilization(U).

### 3.4  DCF Inter Frame Spacing(DIFS) Adaptation

Third scheme that we propose for adapting MAC Layer for providing QoS in WSNs is DIFS Adaptation per class. DIFS, is the time interval since the last sending of frame, after which any node can try to acquire the channel to send a new frame. In proposed framework, we provide intra node service differentiation based on DIFS, for different class of traffic. So, for class I we define parameter denoted by difsI and for class II we define difsII. Values of DIFS for different classes of traffic are, for class I, difsI is 8 and for class II, difsII is 15. Class I has to wait less after sending of last frame to send a new Class I traffic frame, than the Class II frame. In this differentiated service is provided based on MAC layer parameter DIFS. In this case, the DIFS of the SYNC packets has to be adjusted to the DIFS value of Class I, for maintaining the synchronization in the nodes.

### 3.5  Combining all the schemes

Fourth scheme that we propose for adapting MAC Layer for providing QoS in WSNs is implementing all the previous three approaches together, CW Adaptation, Duty Cycle Adaptation, DIFS Adaptation. In this, contention window adaptation is done as specified in Algorithm 1. MAC Delay is measured as specified in Equation 1 and CW is adapted according to the class of traffic to be transmitted. Duty Cycle adaptation also occurs simultaneously, each node sends the information of DSS delay of class I packet to the receiving node, and receiving node keeps on updating the average DSS delay of classI packets in that round, until its time to broadcast the SYNC packets. At this instant it runs the Algorithm 2, to determine the change and inform the nodes. DIFS Adaptation is done before the nodes starts communicating, different DIFS intervals are set for class I traffic and class II traffic for intra node service differentiation. SYNC packets follow DIFS interval of classI packets, as they are important messages for synchronization. In this way all the three proposed schemes works together

---
**Algorithm 2** Procedure for Duty Cycle Adjustment of Class I
---
$U := (T_{rx} + T_{tx}) \div (T_{rx} + T_{tx} + T_{idle})$
{$T_{rx}$ is the receiving time, $T_{tx}$ is transmitting time, $T_{idle}$ is the idle time, in last SYNC period }
if ($U < U_{min}$) then
  $DC := DC_{def}$
  return
end if
{$DC_U$ is calculated duty cycle of node according to its utilization }
if ($U > U_{min}$) then
  $DC_U = \min(DC(1 + (U - Uprev)/Uprev), DC_{max})$
  $DC_U = \max(DC_U, DC_{min})$
  {DC is calculated based on DSS delay }
  if $\rho > \rho_{min}$ then
    $DC := DC^{prev} * (1 + (S - S_I) \div S_I)$
    if $(abs((DC - DC^{prev})/DC^{prev}) < DC\ thresh)$ then
      $DC := DC^{prev}$
      return
    end if
    {if DC based on DSS delay is less than $DC^{prev}$, then DC is based on utilization}
    if ($DC < DC^{prev}$) then
      $DC := \max(DC_U, DC_{min})$
    else
      {if DC based on DSS delay is greater than $DC^{prev}$, then DC is minimum of $DC_U$ and DC based on DSS delay, so to minimize energy consumption}
      if ($DC > DC^{prev}$) then
        $DC := \min(DC, DC_U)$
      end if
    end if
  end if
end if

in a node to improve the end-to-end delay for classI(delay sensitive) traffic in presence of classII(periodic reporting) traffic.

## 4  Cross Layer Extension

### 4.1  Adapting duty cycle based on network layer information

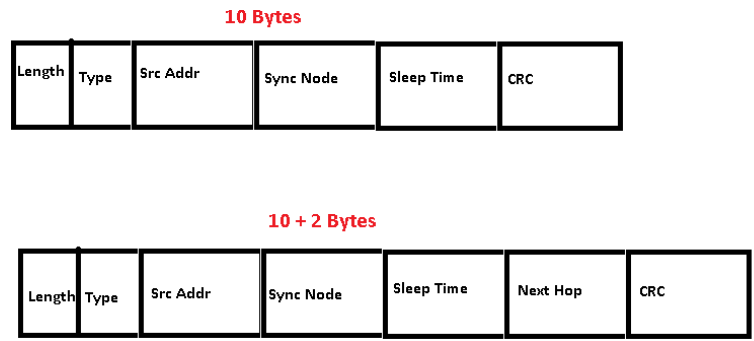

Fig. 1. SYNC packet structure

In SMAC [1], SYNC packets are periodically broadcasted for maintaining their sleep schedules, within this SYNC packet nodes can include their next hop information for the next node towards destination(Base Station). Nodes on receiving these SYNC packets can find out whether they lie in the path from source to destination(Base station). In Figure 4, nodes 1,2,3,4,5 are next hops to more number of nodes. In larger WSN scenarios, this effect would be more prominent. Figure 1 shows the original and modified packet formats.

In Algorithm 3, $N_{nexthop}$ and $N_{min\ nexthop}$ parameters are used to control the unnecessary duty cycle adaptation. In Algorithm 2, duty cycle adaptation is based on Utilization and DSS delay of the node, it can be the case that only few nodes cause high Utilization and DSS delay at their communicating peer, and cause it to change its duty cycle which results in energy inefficient usage in WSNs. Algorithm 3, handles this case by network layer information of next hop. SYNC packets gives the information of how many nodes have this node as the next hop for base station as destination, and based on that node can calculate $N_{nexthop}$. It calculates parameter $N_{nexthop}$ which is the number of different nodes for which this node is selected as next hop in the last SYNC period. Then if it is greater then $N_{min\_nexthop}$ threshold, it applies usual duty cycle adaptation based on DSS delay and Utilization of the node. Node can select optimum value of $N_{min\_nexthop}$, based on its information of all its next hop neighbors.

---

**Algorithm 3** Procedure for Duty Cycle Adjustment of Class I according to next hop information
---

$U := (T_{rx} + T_{tx}) \div (T_{rx} + T_{tx} + T_{idle})$
{$T_{rx}$ is the receiving time, $T_{tx}$ is transmitting time, $T_{idle}$ is the idle time, in last SYNC period }
if $(U < U_{min})$ then
   $DC := DC_{def}$
   return
end if
if $(N_{nexthop} > N_{min\ nexthop})$ then
   {$N_{nexthop}$ is the number of different Src nodes for which this node was selected as next hop, $N_{min\_nexthop}$ is the threshold number of times a node should be elected as next hop to do duty cycle adaptation in last SYNC period}
   {$DC_U$ is calculated duty cycle of node according to its utilization }
   if $(U > U_{min})$ then
     $DC_U = min(DC(1 + (U - Uprev)/Uprev), DC_{max})$
     $DC_U = max(DC_U, DC_{min})$
     {DC is calculated based on DSS delay }
     if $\rho > \rho_{min}$ then
       $DC := DC^{prev} * (1 + (S - S_I) \div S_I)$
       if $(abs((DC - DC^{prev})/DC^{prev}) < DC\ thresh)$ then
          $DC := DC^{prev}$
          return
       end if
       {if DC based on DSS delay is less than $DC^{prev}$, then DC is based on utilization}
       if $(DC < DC^{prev})$ then
          $DC := max(DC_U, DC_{min})$
       else
          {if DC based on DSS delay is greater than $DC^{prev}$, then DC is minimum of $DC_U$ and DC based on DSS delay, so to minimize energy consumption}
          if $(DC > DC^{prev})$ then
             $DC := min(DC, DC_U)$
          end if
       end if
     end if
   end if
end if

### 4.2 Adapting route calculation based on DSS delay parameter of the MAC layer

In Wireless Sensor Networks, the MAC layer protocol SMAC[1], nodes form virtual clusters based on common sleep schedules to reduce control overhead and enable traffic-adaptive wake-up. So when data is sent from source to sink, at many nodes it has to go from one virtual clusters to another virtual cluster, and the bordering node follows both the sleep schedules. This difference in sleep schedule is incorporated in total end-to-end delay. Delay incurred because of two upcoming communicating nodes following different sleep schedules, is termed as DSS delay of the of the node. In our proposed scheme, network layer routing protocol takes into consideration this DSS delay of the link into the cost of the link and computes its overall cost, for route calculation.

**Estimation of DSS delay of link** The measurement technique for Difference in Sleep Schedules Delay for link is very simple and similar to the calculation of average DSS delay in Equation 2. A transmitting node computes this delay($s_{link}$), by subtracting the time ($ts_s$) that a packet is passed to the MAC layer from the time ($ts_c$) it starts carrier sensing for sending the packet. Node then stores this delay as $s^i_{link}$, and calculates the average delay $s^i_{linkavg}$ for that link. Here $s^i_{link}$ and $s^{i-1}_{linkavg}$ are measured DSS delay for that link, and previously stored average DSS delay for that link. $\beta$ is a positive constant, which determines how much effect the previously stored average DSS delay for that link have on the current average DSS delay for that link. $S_I$ is the expected maximum delay, same as used in Algorithm 2.

$$s^i_{linkavg} = (1 - \beta) * s^i_{link} + \beta * s^{i-1}_{linkavg} \qquad (3)$$

---

**Algorithm 4** Procedure for calculating overall link cost including DSS delay of the link

---

{$LC_{overall}$ is calculated based on DSS delay}
if $S_{link} > S_I$ then
　　$LC_{overall} := LC_{ori} * (1 + (S_{linkavg} - S_I) \div S_I)$
else
　　$LC_{overall} := LC_{ori}$
end if

---

In the Figure 2, Node 4 is the sending node towards the base station, if routes 4–>3–>2–>1, and route 4–>5–>6–>7, have same link costs, but different DSS delay of constituting links, then route 4–>3–>2–>1, with lower DSS delays of the constituting links will give less end-to-end latencies, compared to other routes. Our overall link cost $LC_{overall}$ in Algorithm 4, takes care of such cases ,and facilitates the selection least end-to-end latency routes for delay sensitive traffic in WSNs.

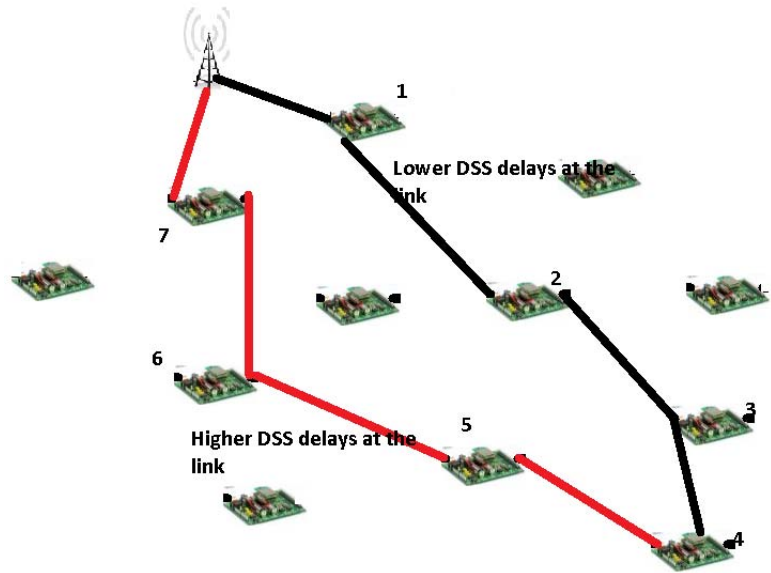

Fig. 2.  Route Adaptation

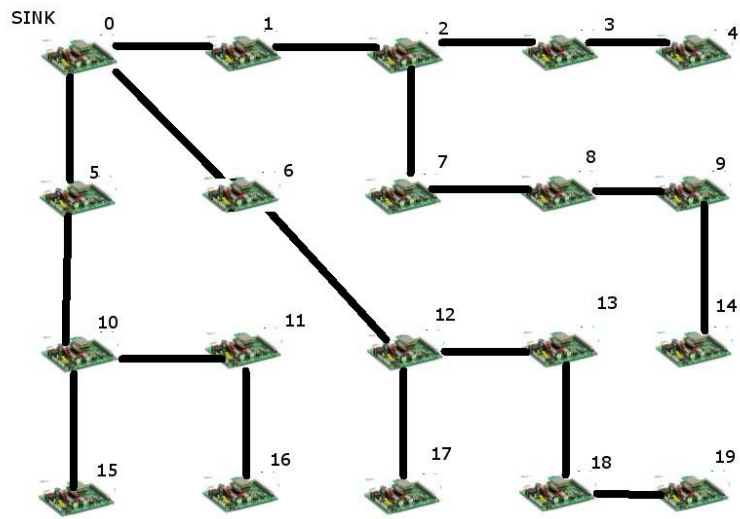

Fig. 3.  Scenario 1

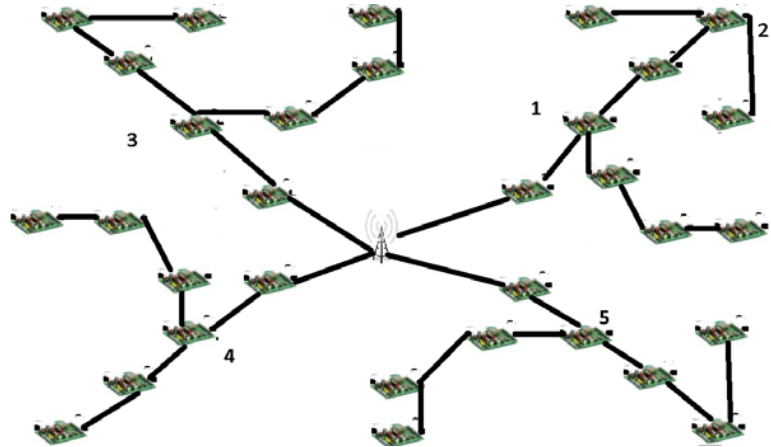

Fig. 4. Scenario 2

## 5  Simulation And Results

In this section simulations of the MAC layer adaptation for QoS in WSN are discussed and analyzed over two simulation scenarios with different topological and traffic conditions. The simulation of proposed MAC layer adaptation scheme is implemented in ns2.29 simulator[10]. One topology that we used for our simulations is shown in Figure 3. In Figure 3, the Node numbered 0 is sink and all other sensors transmit to this node. Equal number of Class I and Class II packets are generated by each node. In this topology, distance between two nodes horizontally and vertically is 140 meters, and diagonally its around 196 meters. Communication range of each node is around 200 meters. Lines shown in the Figure 3, are routes. For simplicity, we have used static routing with only one next hop from each node, but our approach is not constrained by the topology.

Figure 5, shows the Contention Window Adaptation, Figure 7, Duty Cycle Adaptation, Figure 6, shows the DIFS Adaptation and Figure 8, shows the combination of all three proposals together. In every graph, the X axis is the Number of packets of each class received at the sink, and the Y axis is the cumulative end to end delay of packets of each class received at the sink. Since the number of packets received at sink in SMAC and Proposed scheme are same, average latency of two schemes are reported for all traffic.

In Figure 5, Contention window adaptation is done based on per class basis as proposed in Section 3.2. After the adaptation delay of classII packets went greater than original SMAC classII packets, and delay of classI packets became less than original classII SMAC traffic. This is because of per class differentiation done in the Algorithm 1, which gives priority to classI traffic(delay sensitive) than classII(best effort) traffic. Contention Window Adaptation, leads to 30% benefit in average end-to-end delay of Class I Packets compared to SMAC.

In Figure 7, duty cycle adaptation is shown as proposed in section 3.3. In Figure 7, average latency of classI and calssII traffic both reduces compared to original SMAC, as increasing duty cycle would reduce the latencies for both class of traffic. ClassI average latency is lesser than classII average latency, on duty cycle adaptation. Duty Cycle Adaptation, leads to 37% benefit in average end-to-end delay of Class I Packets compared to SMAC.

In Figure 6, per level differentiation is done on the basis of DIFS values. After the adaptation delay of classII packets went greater than original SMAC classII packets, and delay of classI packets became less than original classII SMAC traffic. This is because of per class differentiation done, which gives priority to classI traffic(delay sensitive) than classII(best effort) traffic. DIFS Adaptation, leads to 25% benefit in average end-to-end delay of Class I Packets compared to SMAC.

In Figure 8, all the three adaptation are shown simultaneously, Duty Cycle, CW, DIFS Adaptation combined leads to 60% benefit in average end-to-end delay of Class I Packets compared to SMAC.

Second topology that is used for analysis is shown in Figure 4. This comprises of more number of sensors than in Topology 3 and different traffic conditions. In this topology horizontal and vertical nodes are separated by 250 meters, and communication range of each node is around 300 meters. Nodes generate equal amount of classI and classII traffic. In Figure 9, this topology is evaluated with all the three MAC layer adaptations and leads to around 55% benefit in average end-to-end delay of Class I Packets compared to SMAC.

## 6   CONCLUSION

The paper, proposed schemes to deal with delay sensitive(event driven) traffic in presence of periodic reporting traffic. This is achieved by cross layer adaptation involving MAC layer and Network layer. MAC layer adaptation is achieved by adapting (i) CW, different contention window for different class of traffic (ii) Duty Cycle, adapted appropriately according to Utilization and DSS delay,unlike the previous works in literature (iii) DIFS parameters at MAC layer (iv) Combination of all three schemes proposed above. In dynamic duty cycle adaptation, our scheme adapts duty cycle by appropriate amount instead of doubling or halving it as done in previous approaches[2], so our scheme reduces energy consumption for duty cycle adaptation, but at cost of extra control messages overhead. The simulation results establish superiority of all the proposed schemes over SMAC. All the four proposed schemes shows the improvement in end-to-end delay of class I traffic in presence of class II traffic. In cross layer adaption, two mechanisms are proposed, (i) MAC layer utilizing next hop information from incoming SYNC packets to judiciously adapt the duty cycle of the node. (ii) Network layer utilizing MAC layer parameter DSS delay of the link, into consideration while computing routes, so as to select the best routes suitable for delay sensitive traffic.

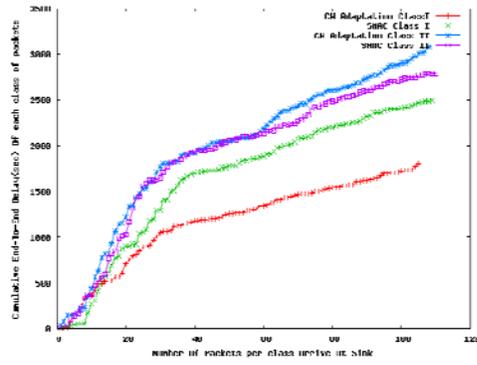 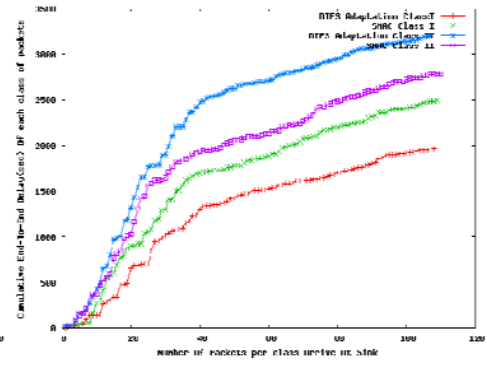

Fig. 5. Cumulative end-to-end delay of packets at sink with, CW adaptation

Fig. 6. Cumulative end-to-end delay of packets at sink, with DIFS Adaptation

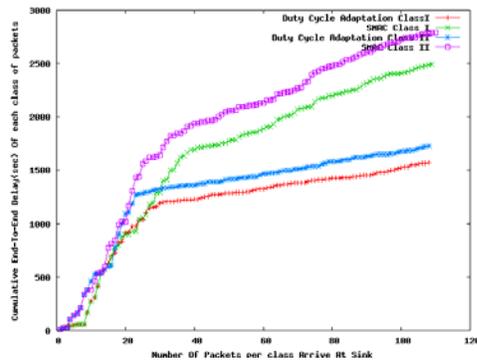 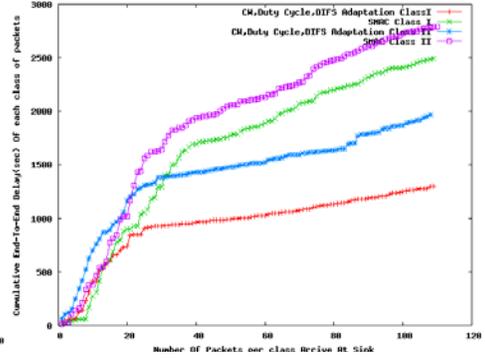

Fig. 7. Cumulative end-to-end delay of packets at sink, with Duty Cycle Adaptation

Fig. 8. Cumulative end-to-end delay of packets at sink, with all three adaptations(CW,Duty Cycle,DIFS)

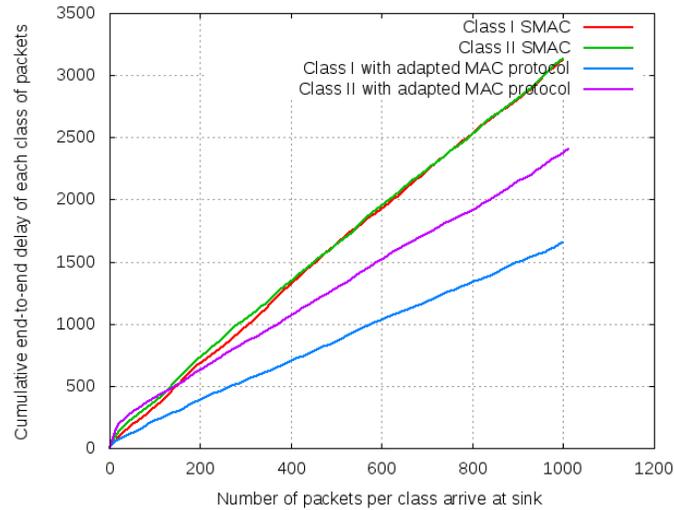

Fig. 9. Cumulative end-to-end delay of packets at sink, with all three adaptations(CW,Duty Cycle,DIFS) using Scenario 2